\documentclass[useAMS,usenatbib]{mn2e}
\usepackage{natbib}
\usepackage{pbox}
\usepackage{aas_macros}
\usepackage{amsmath,amssymb}
\usepackage{booktabs}
\usepackage{tabularx}
\newcommand*{\rom}[1]{\expandafter\@slowromancap\romannumeral #1@}

\usepackage{xfrac}
\usepackage{pgfplots}
\usepackage{cleveref}

\crefname{figure}{Fig.}{Figs}

\usepackage{pgfplots}
\usepackage{amssymb}
\usepackage{gensymb}

\pgfplotsset{compat=1.5}

\title[Analysis of g modes in KIC\,10080943]{KIC\,10080943: a binary star
  with two $\gamma$\,Doradus/$\delta$\,Scuti hybrid pulsators. Analysis
  of the g modes} 
\author[M. A. Keen {\em et. al.}] 
{M. A. Keen$^{1,2}$\thanks{E-mail: mkee4218@uni.sydney.edu.au (MAK)}, 
	T. R. Bedding$^{1,2}$\thanks{E-mail: bedding@physics.usyd.edu.au (TRB)},
	S. J. Murphy$^{1,2}$\thanks{E-mail: murphy@physics.usyd.edu.au (SJM)},
	V. S. Schmid$^{3}$,
	C. Aerts$^{3,4}$, \and
	A. Tkachenko$^{3}$,
	R.-M. Ouazzani$^{2}$ and 
	D.W. Kurtz$^{5}$
\\ 
$^{1}$Sydney Institute for Astronomy, School of Physics, The University of Sydney, NSW 2006, Australia\\
$^{2}$Stellar Astrophysics Centre, Department of Physics and Astronomy, Aarhus University, Denmark\\
$^{3}$Institute of Astronomy, KU Leuven, Celestijnenlaan 200D, B –- 3001 Leuven, Belgium\\
$^{4}$Department of Astrophysics, Radboud University Nijmegen, P.O. Box 9010, 6500 GL Nijmegen, The Netherlands\\
$^{5}$Jeremiah Horrocks Institute, University of Central Lancashire, Preston PR1 2HE, UK\\
}

\begin{document}

\pagerange{\pageref{firstpage}--\pageref{lastpage}} \pubyear{2015}

\maketitle

\label{firstpage}


\begin{abstract}
We use four years of {\em Kepler} photometry to study the non-eclipsing
spectroscopic binary KIC\,10080943.  We find both components to be
$\gamma$\,Doradus/$\delta$\,Scuti hybrids, which pulsate in both p and g
modes.  We present an analysis of the g modes, which is complicated by the
fact that the two sets of $\ell=1$ modes partially overlap in the frequency
spectrum.  Nevertheless, it is possible to disentangle them by identifying
rotationally split doublets from one component and triplets from the other.
The identification is helped by the presence of additive combination
frequencies in the spectrum that involve the doublets but not the triplets.
The rotational splittings of the multiplets imply core rotation periods of
about 11\,d and 7\,d in the two stars.  One of the stars also shows
evidence of $\ell=2$ modes.
\end{abstract}



\section{Introduction}

Pulsating stars in binary systems have great potential to contribute to our
understanding of stellar astrophysics. The orbit provides mass constraints
for the two stars, while modelling of the pulsations (asteroseismology)
yields information about their internal structure and rotation. When both
stars are pulsating, such binary systems are particularly valuable.

Stars can oscillate in both pressure modes (p\:modes), which are most
sensitive to conditions in the outer layers of the star, and gravity modes
(g\:modes), which are most sensitive to the interior
\citep[e.g.,][]{Aerts2010}.  Observations with the {\em Kepler} spacecraft
have revealed that many main-sequence A and F stars show both
$\delta$\,Scuti (p-mode) and $\gamma$\,Doradus (g-mode) pulsations
\citep{Grigahcene2010a,Uytterhoeven2011,Tkachenko2013,Balona2014}.

Detailed studies of $\delta$\,Sct/$\gamma$\,Dor hybrids have so far been
carried out for two stars, namely KIC\,11145123 \citep{Kurtz2014} and
KIC\,9244992 \citep{Saio2015}.  Both of those stars rotate slowly (periods
of $100\,\textrm{d}$ and $66\,\textrm{d}$, respectively), with the rotation
of the core differing only slightly from the rotation at the surface. The
results showed that a strong angular momentum transport mechanism must be
at work in main-sequence, intermediate-mass stars.

Here we discuss KIC\,10080943, which is a non-eclipsing, double-lined
spectroscopic binary \citep{Tkachenko2013} in which we find both components
to be $\gamma$\,Dor/$\delta$\,Sct hybrids with moderate rotation.
\cite{Huber2014} listed the following properties for KIC\,10080943:
\mbox{$T_{\textrm{eff}} = 7360\pm260 \, \textrm{K}$}, \mbox{$\log g =
4.0\pm0.4$} and $\mbox{[Fe/H]} = -0.2 \pm 0.3$, while the visual magnitude
was measured by \cite{Tkachenko2013} to be \mbox{$V = 11.7 $}.

This paper presents our analysis of the g-mode spectrum of KIC\,10080943.
A companion paper by \citet{schmidetal2015} discusses the p modes and the
binary properties, as well as addressing the individual stellar parameters of the
constituent stars, as has been done for eclipsing binaries with intermediate-mass
primaries \citep{Hambleton2013,Debosscher2013,Maceroni2014}. In particular,
we have measured phase modulations in this system
\citep{murphyetal2014,murphy&shibahashi2015}, showing that orbital variations
in the light arrival times of some of the p-mode pulsations occur in
anti-phase with others. This demonstrates that both stars pulsate in
p\:modes as discussed by \citet{schmidetal2015}. The same methodology
cannot be applied to the g\:modes because the higher frequency density demands
an unachievable frequency resolution, given the sampling required for the short orbit. Nonetheless, the focus of this paper is to show that both components also pulsate in g\:modes.


\begin{figure*}
\begin{center}
	\includegraphics[width=0.98\textwidth]{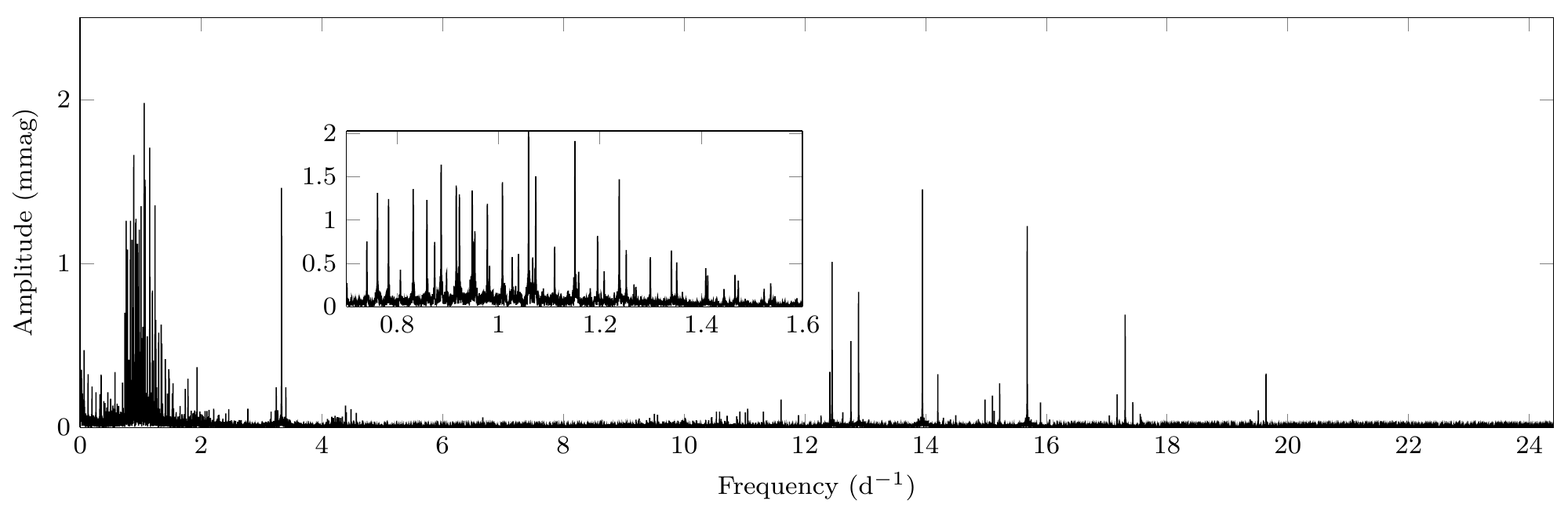}
	\caption{Fourier amplitude spectrum of KIC\,10080943 from {\em
	Kepler} long cadence data, showing the clearly separated g-mode and
	p-mode regions. Inset: close-up of the g\:modes.}
	\label{fig:FourierSpectrum}
\end{center}
\end{figure*}

\begin{figure*}
\begin{center}
\includegraphics[width=0.98\textwidth]{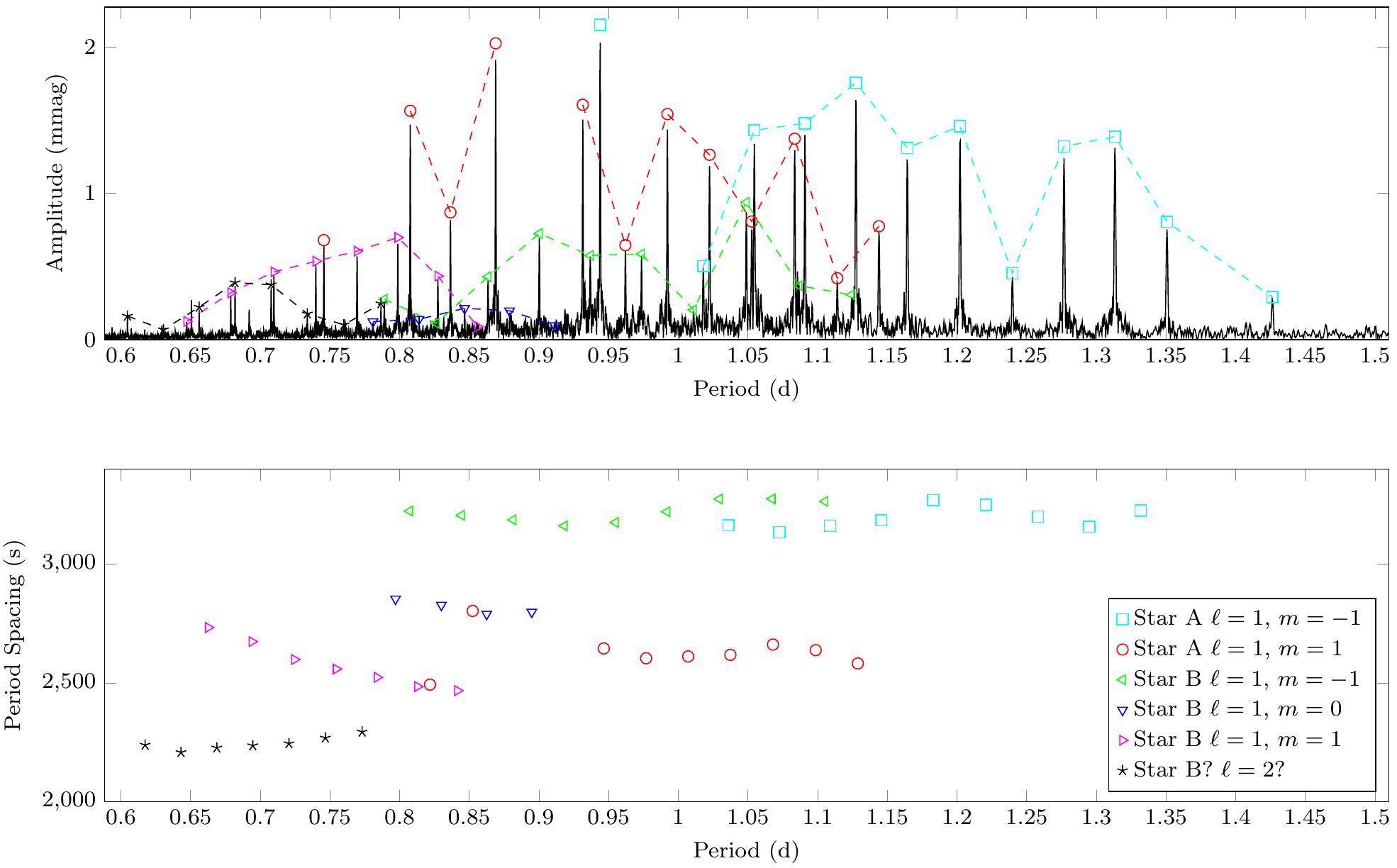}
\caption{Top: Amplitude spectrum of the g-mode region, plotted as a
function of period, showing the identified peaks. Dashed lines connect neighbouring peaks in the same g-mode series. Bottom: Period
spacings between adjacent modes in each series. }
\label{fig:Polygon}
\end{center}
\end{figure*}

\begin{figure*}
\begin{center}
\includegraphics[width=0.98\textwidth]{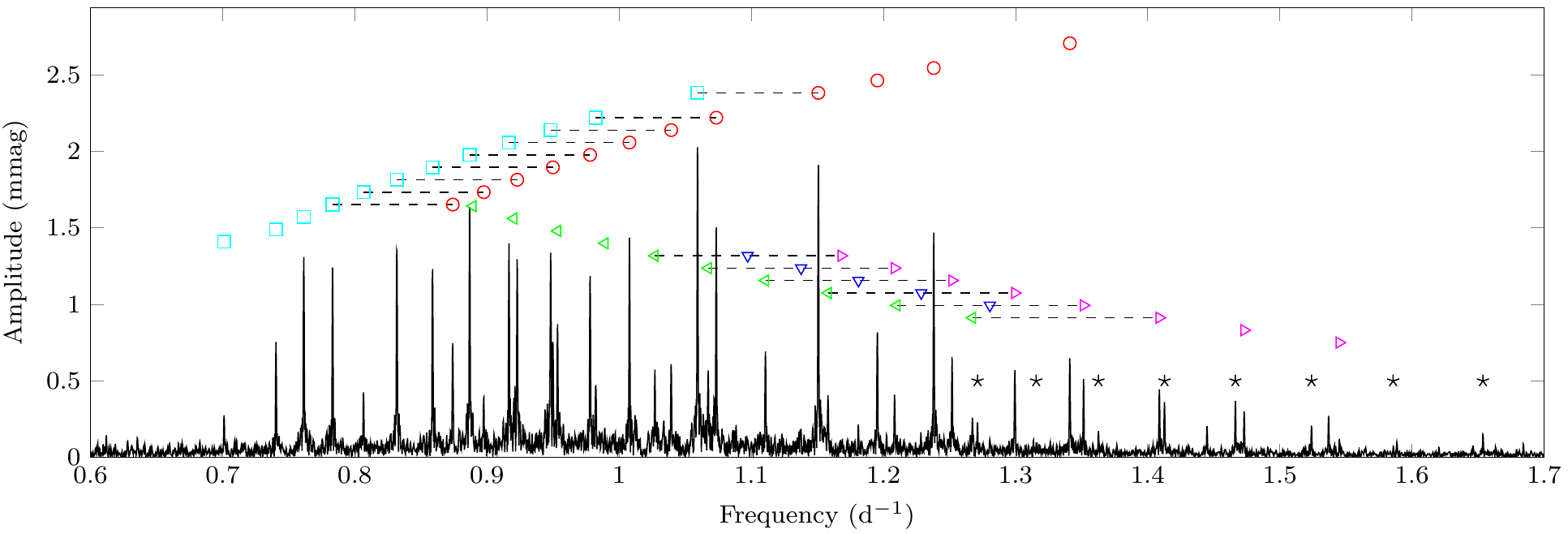}
\caption{Identification of rotationally split g-mode multiplets in KIC\,10080943. Dashed lines connect peaks in each multiplet. Colours and symbols used match those of \Cref{fig:Polygon}.}
\label{fig:Triplet}
\end{center}
\end{figure*}

\begin{figure}
\begin{center}
   \includegraphics[width=0.45\textwidth]{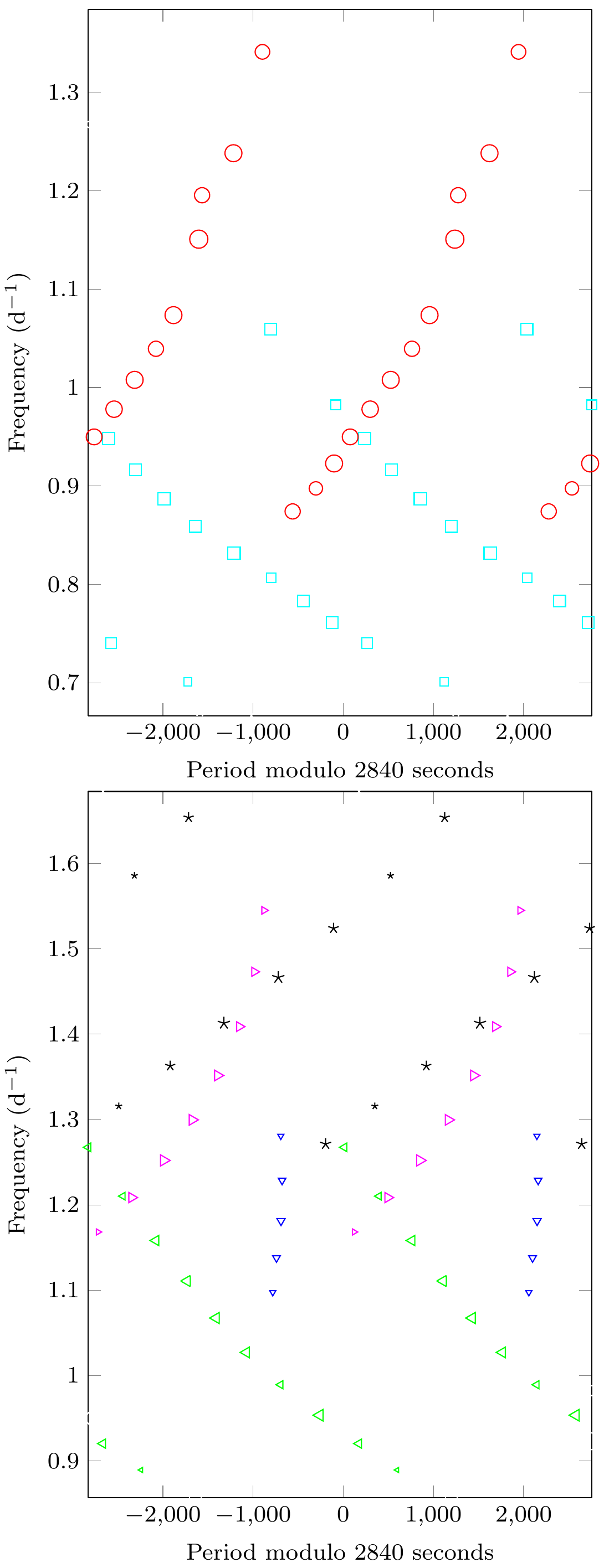}
   \caption{Period {\'e}chelle diagrams of the g\:modes in KIC\,10080943,
   for Star~A (upper) and Star~B (lower).  The data have been plotted twice
   for clarity.  The symbol size is proportional to the logarithm of the
   amplitude of the corresponding peak in the Fourier spectrum.}
   \label{fig:Echelle}
\end{center}
\end{figure}

\section{Data Analysis \& Results}

When analysing high-order g modes, we are guided by the expectation that
they are equally spaced in period \citep{Tassoul1980}. However, the steep
composition gradient ($\mu$ gradient) between the core and outer layers can
cause deviations from this uniform period spacing
\citep{Miglio2008}. Furthermore, as shown by \cite{Bouabid2013}, stellar
rotation causes the period spacing to vary with frequency. These effects
have been observed in {\em Kepler} data for a number of $\gamma$\,Dor stars
\citep{Bedding2014,VanReeth2015a,VanReeth2015b}.

For KIC\,10080943, we used long-cadence data (29.4-minute sampling) from the
full {\em Kepler} mission, spanning 1470.5\,d (4.0\,y). The
Fourier spectrum (\Cref{fig:FourierSpectrum}) shows pulsations in both the
g-mode and p-mode frequency regions. These regions are clearly separated,
with the g-mode region below approximately 6\,d$^{-1}$ and the p-mode
region above approximately 8\,d$^{-1}$.

At low frequency there is a peak at 0.065209\,d$^{-1}$ with a long series of harmonics. Such harmonic series are often seen in binary stars. The inverse of this frequency is 15.335\,d, which we take as the binary orbital period. KIC\,10080943 was already established as a binary by \citet{Tkachenko2013}, and was selected for spectroscopic observations to determine the orbital parameters of the system. Those are presented in \citet{schmidetal2015}.

To analyse the pulsation content, we sequentially extracted
the 250 strongest peaks in the frequency range
from zero to the Nyquist frequency (24.47\,d$^{-1}$) using {\small
PERIOD04} \citep{Lenz2004,Lenz2005}. Non-linear least-squares fitting was
implemented during extraction. Although we were interested in the
g\:modes, the p\:modes were also extracted because the spectral windows of
the p and g\:modes overlap. We also checked that the extracted peaks
were not Nyquist aliases using the method outlined by
\cite{Murphy2013}.

In the range 0.6 to 1.7\,d$^{-1}$ we identified six series of pulsation
modes. Each is approximately equally spaced in period, consistent with a
series of overtones with a common $\ell$ and $m$ value. Furthermore, some
of the series are separated from others by a fixed frequency, suggesting
rotational splitting.  Given that KIC\,10080943 is a binary system,
 we were led to identify two sets of g modes
with slightly different period spacings, one from each component of the
binary. This is shown in Figs\,\ref{fig:Polygon} and \ref{fig:Triplet}, where
we have arbitrarily labelled the components of the binary as Stars~A and~B.
In Star~A we see rotationally split doublets, while Star~B has triplets in
which the central component is weak. There is also a sequence with slightly
lower period spacing (asterisks) that we tentatively identify with $\ell=2$
modes.  Note that \cite{Bedding2014} have identified clear evidence for
$\ell=2$ g\:modes in another $\gamma$~Doradus star, KIC\,3127996.

The peaks are plotted in {\'e}chelle format in \Cref{fig:Echelle} for the
two stars, with symbol size indicating amplitude.  The frequencies and
amplitudes for the six g-mode series are given in
\Cref{tab:doublet,tab:triplet,tab:singlet}.  Importantly, all significant
peaks in this frequency range have been identified with modes.
The number of points in the {\'e}chelle diagram (\Cref{fig:Echelle}) is greater
than the number in the period spacing diagram (Fig.\,\ref{fig:Polygon}), since the
latter require two neighbouring modes for each period spacing, but some modes
are missing. Values of $m$ could still be assigned to modes in incomplete
multiplets by tracing their period spacings.

\subsection{Combination Frequencies}

\citet{kurtzetal2015} showed recently that the Fourier spectra of
some g-mode pulsators are dominated by combination frequencies.  We have
checked whether any peaks in the main g-mode region of KIC\,10080943 (0.6
to 1.7\,d$^{-1}$) are combinations of other peaks.  This was done by
creating linear combinations (including harmonics) of the 15 strongest
peaks, using up to three of these peaks with integer coefficients from $-3$
to~$3$.  None of the observed frequencies were found to be within
$1/T=6.8\times10^{-4}$\,d$^{-1}$ of a calculated combination frequency.
This is not surprising, given that all these peaks have already been
identified with one of the six series of overtones.

\begin{figure*}
\begin{center}
\includegraphics[width=0.95\textwidth]{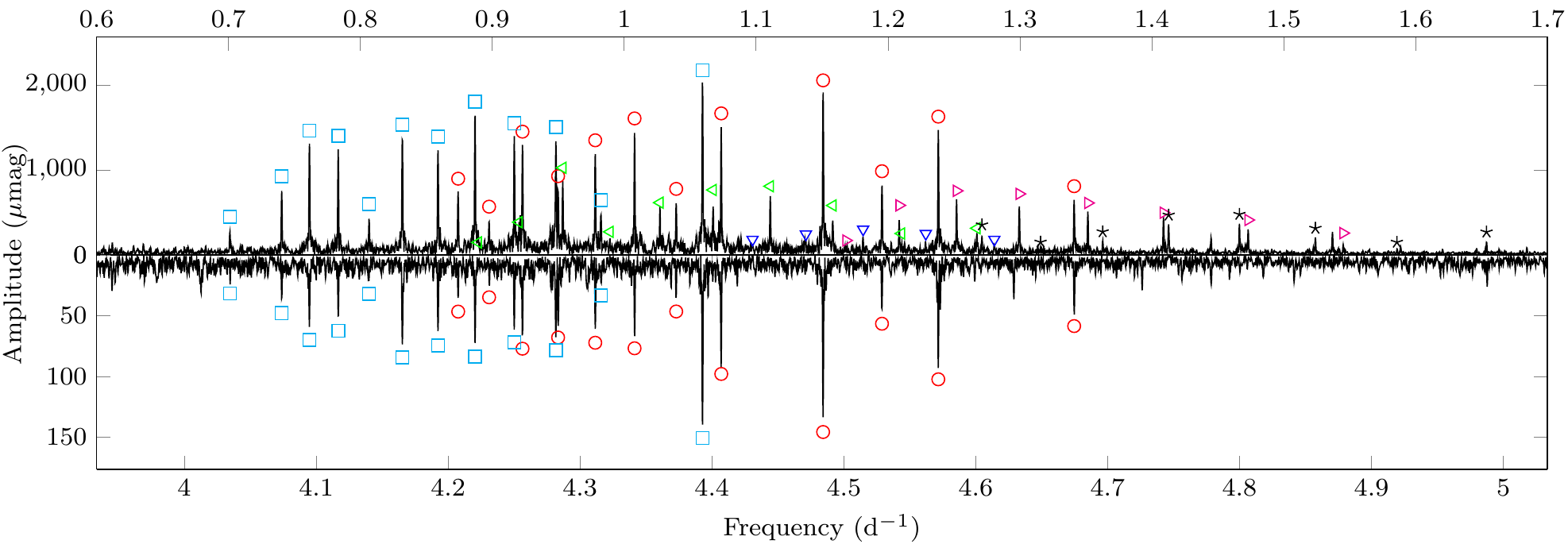}
\caption{Additive combination frequencies in KIC\,10080943.  The upper plot
  shows the g-mode region of the amplitude spectrum while the lower plot
  shows a higher-frequency section of exactly the same length, with an
  offset of \mbox{$f_{3.3} = 3.33 \, \textrm{d}^{-1}$}.  Coloured symbols have
  the same meanings as in previous figures, with the addition that those in
  the lower panel indicate combination frequencies.  Note the different
  amplitude scales in the two panels.}
\label{fig:HumpUpsideDown}
\end{center}
\end{figure*}

On the other hand, the small hump in the frequency range 4.0 to
4.8\,d$^{-1}$ is made up of combination peaks.  To be specific, these peaks
have frequencies $f_{3.3} + f_g$, where \mbox{$f_{3.3} = 3.3334962 \, \pm
  0.0000017 \, \textrm{d}^{-1}$} is one of the strongest peaks in the
amplitude spectrum.  As can be seen in \Cref{fig:HumpUpsideDown}, the doublets
we have assigned to Star~A couple with $f_{3.3}$ to produce additive
combination frequencies, but the triplets we have assigned to Star~B do not.
This suggests that $f_{3.3}$ originates in Star~A and is further evidence
that g-mode pulsations occur in both components of the binary. Subtractive ($f_{3.3} - f_g$) combinations are also present.

\subsection{Rotational Splittings}

\begin{figure}
\begin{center}
\includegraphics[width=0.48\textwidth]{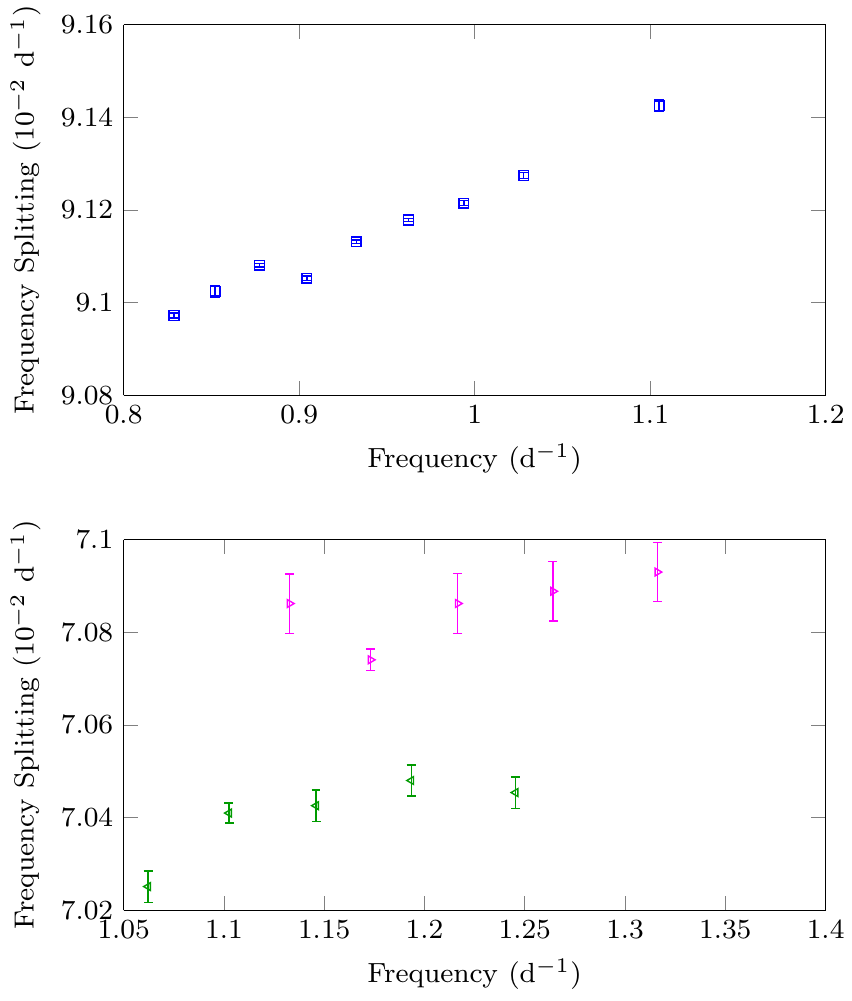}
\caption{Upper panel: rotational splittings of the doublets in Star A,
measured between the $m=-1$ and $m=1$ modes for each radial order.  Lower
panel: rotational splittings of the triplets for Star B. The upper sequence
(magenta symbols) show the splittings measured between $m=0$ and $m=1$,
while the lower sequence (green symbols) show the splittings between $m=-1$
and $m=0$.}
\label{fig:FreqSplitting}
\end{center}
\end{figure}

As discussed above, we identified two series of rotationally split $\ell=1$
multiplets, one in each component of the binary -- see \Cref{fig:Triplet}.
The triplets in Star~B ($m=0, \pm1$) have frequency splittings of about
0.07\,d$^{-1}$ and are slightly asymmetric, as shown in the lower panel of
\Cref{fig:FreqSplitting}. The asymmetry is small, at about 0.5\:per\:cent,
and is presumably caused by second-order effects (see \citealt{murphy2014}
for a discussion of non-equal rotational splittings in A stars in the \textit{Kepler} era).
The implied rotation period at the edge of the convective core is about 7\,d.

The doublets in Star~A, which presumably correspond to azimuthal orders
$m=\pm1$, have a frequency splitting of 0.091\,d$^{-1}$.  This varies
slightly with frequency, as shown in the upper panel of
\Cref{fig:FreqSplitting}.  This variation cannot be explained by radial
differential rotation because the g-modes all probe the same region outside the core.
Instead, it reflects the fact that the Ledoux constant $C_{n,l}$ has a small dependence
on radial order, $n$ (Bedding et al., in prep.). 
Star~A is a particularly good example of this phenomenon.  Its core rotation period is about 11\,d.

\section{Conclusion}

KIC\,10080943 is a binary system with two pulsating components, making it
particularly interesting for asteroseismology.  The combined oscillation
spectrum is complicated because the two sets of g modes partially
overlap in frequency.  However, it was possible to disentangle them by
identifying rotationally split doublets in Star~A and triplets in Star~B.
The identification was helped by the presence of additive combination
frequencies in Star~A and not Star~B.  The rotational splittings of the
g-mode multiplets imply core rotation periods of about 11\,d and 7\,d in
Stars~A and~B, respectively.  Analysis of the p modes, together with models
of the oscillation frequencies, are presented by \citet{schmidetal2015} and
in future papers.


\section{Acknowledgements}

We gratefully acknowledge the entire {\em Kepler} team for providing such
superb data.  This research was supported by the Australian Research
Council. Funding for the Stellar Astrophysics Centre is provided by the
Danish National Research Foundation (grant agreement no.:DNRF106). The
research is supported by the ASTERISK project (ASTERoseismic Investigations
with SONG and {\em Kepler}) funded by the European Research Council (grant
agreement no.: 267864). VAS is an Aspirant PhD Fellow and AT is a Postdoctoral
Fellow of the Fund for Scientific Research of Flanders (FWO), Belgium.

\begin{table}
		\caption{A least-squares fit of the g-mode $\ell = 1$ doublets in Star\:A (as shown in \Cref{fig:Triplet}) for KIC\,10080943. The uncertainties were determined by non-linear least-squares fitting.} 
	\centering
	\begin{tabular}{rcc}
		\toprule 
		m & Frequency & Amplitude \\
		 & ($\textrm{d}^{-1}$) & (mmag) $\pm$ 0.007 \\
		\midrule 
		$-1$ & $0.701089\; \pm \;0.000011$ & $0.237$ \\
		\midrule
		$-1$ & $0.740343\; \pm \;0.000003$ & $0.726$ \\
		\midrule
		$-1$ & $0.761387\; \pm \;0.000002$ & $1.314$ \\
		\midrule
		$-1$ & $0.783179\; \pm \;0.000002$ & $1.163$ \\
		{ }1 & $0.874151\; \pm \;0.000003$ & $0.762$ \\
		\midrule
		$-1$ & $0.806578\; \pm \;0.000006$ & $0.440$ \\
		{ }1 & $0.897602\; \pm \;0.000006$ & $0.415$ \\
		\midrule
		$-1$ & $0.831815\; \pm \;0.000002$ & $1.414$ \\
		{ }1 & $0.922896\; \pm \;0.000002$ & $1.436$ \\
		\midrule
		$-1$ & $0.858849\; \pm \;0.000002$ & $1.263$ \\
		{ }1 & $0.949902\; \pm \;0.000003$ & $0.971$ \\
		\midrule
		$-1$ & $0.886926\; \pm \;0.000002$ & $1.679$ \\
		{ }1 & $0.978058\; \pm \;0.000002$ & $1.239$ \\
		\midrule
		$-1$ & $0.916677\; \pm \;0.000002$ & $1.401$ \\
		{ }1 & $1.007856\; \pm \;0.000002$ & $1.379$ \\
		\midrule
		$-1$ & $0.948214\; \pm \;0.000002$ & $1.474$ \\
		{ }1 & $1.039429\; \pm \;0.000003$ & $0.731$ \\
		\midrule
		$-1$ & $0.982322\; \pm \;0.000005$ & $0.511$ \\
		{ }1 & $1.073596\; \pm \;0.000002$ & $1.493$ \\
		\midrule
		$-1$ & $1.059381\; \pm \;0.000012$ & $2.169$ \\
		{ }1 & $1.150806\; \pm \;0.000001$ & $1.956$ \\
		\midrule
		$-1$ & $1.110715\; \pm \;0.000004$ & $0.709$ \\
		{ }1 & $1.195443\; \pm \;0.000003$ & $0.815$ \\
		\midrule
		{ }1 & $1.238138\; \pm \;0.000002$ & $1.455$ \\
		\midrule
		{ }1 & $1.341122\; \pm \;0.000004$ & $0.671$ \\
		\bottomrule 
	\end{tabular}
	\label{tab:doublet} 
\end{table}

\begin{table}
	\caption{A least-squares fit of the g-mode $\ell = 1$ triplets in Star\:B (as shown in \Cref{fig:Triplet}) for KIC\,10080943. The uncertainties were determined by non-linear least-squares fitting.} 
	\centering
	\begin{tabular}{rcc}
		\toprule
		m & Frequency & Amplitude \\
		 & ($\textrm{d}^{-1}$) & (mmag) $\pm$ 0.007 \\
		\midrule
		$-1$ & $0.889276\; \pm \;0.000051$ & $0.048$ \\
		\midrule
		$-1$ & $0.920191\; \pm \;0.000009$ & $0.286$ \\
		\midrule
		$-1$ & $0.953444\; \pm \;0.000003$ & $0.926$ \\
		\midrule
		$-1$ & $0.989185\; \pm \;0.000015$ & $0.173$ \\
		\midrule
		$-1$ & $1.027054\; \pm \;0.000005$ & $0.516$ \\
		{ }0 & $1.097305\; \pm \;0.000029$ & $0.082$ \\
		{ }1  & $1.168167\; \pm \;0.000035$ & $0.071$ \\
		\midrule
		$-1$ & $1.067338\; \pm \;0.000004$ & $0.665$ \\
		{ }0 & $1.137748\; \pm \;0.000018$ & $0.145$ \\
		{ }1  & $1.208489\; \pm \;0.000005$ & $0.484$ \\
		\midrule
		$-1$ & $1.110715\; \pm \;0.000004$ & $0.709$ \\
		{ }0 & $1.181141\; \pm \;0.000013$ & $0.202$ \\
		{ }1  & $1.252003\; \pm \;0.000004$ & $0.655$ \\
		\midrule
		$-1$ & $1.158160\; \pm \;0.000005$ & $0.483$ \\
		{ }0 & $1.228640\; \pm \;0.000016$ & $0.151$ \\
		{ }1  & $1.299529\; \pm \;0.000004$ & $0.619$ \\
		\midrule
		$-1$ & $1.210161\; \pm \;0.000017$ & $0.153$ \\
		{ }0 & $1.280615\; \pm \;0.000030$ & $0.082$ \\
		{ }1  & $1.351545\; \pm \;0.000005$ & $0.512$ \\
		\midrule
		$-1$ & $1.267386\; \pm \;0.000012$ & $0.215$ \\
		{ }1  & $1.408809\; \pm \;0.000006$ & $0.398$ \\
		\midrule
		{ }1  & $1.473037\; \pm \;0.000008$ & $0.312$ \\
		\midrule
		{ }1  & $1.545034\; \pm \;0.000016$ & $0.158$ \\
		\bottomrule
	\end{tabular}
	\label{tab:triplet} 
\end{table}

\begin{table}
		\caption{A least-squares fit of the series of g-modes of KIC\,10080943 marked with black asterisks in \Cref{fig:Echelle,fig:Triplet,fig:Polygon}. These peaks are tentatively labelled as having $\ell = 2$ and belonging to Star\:B. The uncertainties were determined by non-linear least-squares fitting.} 
	\centering
	\begin{tabular}{cc}
		\toprule
		Frequency & Amplitude \\
		($\textrm{d}^{-1}$) & (mmag) $\pm$ 0.007 \\
		\midrule 
		$1.271255\; \pm \;0.000010$ & $0.254$ \\
		$1.315665\; \pm \;0.000050$ & $0.051$ \\	    
		$1.362751\; \pm \;0.000015$ & $0.173$ \\	    
		$1.412779\; \pm \;0.000007$ & $0.369$ \\		    
		$1.466406\; \pm \;0.000007$ & $0.378$ \\		    
		$1.524018\; \pm \;0.000012$ & $0.214$ \\		    
		$1.585784\; \pm \;0.000050$ & $0.050$ \\	
		$1.653745\; \pm \;0.000015$ & $0.170$ \\
		\bottomrule
	\end{tabular}
	\label{tab:singlet} 
\end{table}

\bibliography{kic10080943}

\bsp

\label{lastpage}

\end{document}